\documentstyle[psfig,epsf,aaspptwo]{article}

\def\kms{\relax \ifmmode {\,\rm km\,s}^{-1}\else \,km\,s$^{-1}$\fi}
\def\ha{\relax \ifmmode {\rm H}\alpha\else H$\alpha$\fi}
\def\hb{\relax \ifmmode {\rm H}\beta\else H$\beta$\fi}
\def\hi{\relax \ifmmode {\rm H\,{\sc i}}\else H\,{\sc i}\fi}
\def\hii{\relax \ifmmode {\rm H\,{\sc ii}}\else H\,{\sc ii}\fi}
\def\h2{\relax \ifmmode {\rm H}_2\else H$_2$\fi}
\def\lha{\relax \ifmmode L_{{\rm H}\alpha}\else $L_{{\rm H}\alpha}$\fi}
\def\shi{\relax \ifmmode \sigma_{{\rm HI}}\else $\sigma_{\rm HI}$\fi}
\def\sh2{\relax \ifmmode \sigma_{{\rm H}_2}\else $\sigma_{{\rm H}_2}$\fi}
\def\degr{\hbox{$^\circ$}}
\def\arcmin{\hbox{$^\prime$}}
\def\arcsec{\hbox{$^{\prime\prime}$}}
\def\deg{\hbox{$^\circ$}}

\def\sec{\hbox{$^{\prime\prime}$}}
\def\fdg{\hbox{$.\!\!^\circ$}}
\def\fs{\hbox{$.\!\!^{\rm s}$}}
\def\farcm{\hbox{$.\mkern-4mu^\prime$}}
\def\farcs{\hbox{$.\!\!^{\prime\prime}$}}
\def\degd#1.#2{ #1\fdg#2 }                 % degrees over decimal point
\def\mind#1.#2{ #1\farcm#2 }               % minutes over decimal point
\def\secd#1.#2{ #1\farcs#2 }               % seconds over decimal point
\def\hhh{\ifmmode {\rm ^h}              % hours symbol
         \else {${\rm ^h}$}
         \fi}
\def\sss{\ifmmode {\rm ^s}              % seconds symbol
         \else {${\rm ^s}$}
         \fi}
\def\hms#1h#2m#3s{                      % hms format (for RA)
                  \relax
                  \ifmmode #1^{\rm h}\,#2^{\rm m}\,#3^{\rm s}
                  \else \hbox{$#1^{\rm h}\,#2^{\rm m}\,#3^{\rm s}$}
                  \fi
                 }
\def\dms#1d#2m#3s{                      % dms format (for Dec)
                  \relax
                  #1\degr\,#2\arcmin\,#3\arcsec 
                 }
\def\hmsd#1h#2m#3.#4s{                  % hms format with decimal point (RA)
                      \relax
                      \ifmmode #1^{\rm h}\,#2^{\rm m}\,#3\fs#4
                      \else \hbox{$#1^{\rm h}\,#2^{\rm m}\,#3\fs#4$}
                      \fi
                     }
\def\dmsd#1d#2m#3.#4s{                  % dms format with decimal point (Dec)
                      \relax
                      #1\degr\,#2\arcmin\,#3\farcs#4
                     }
\def\mag{\relax                          % magnitudes symbol
        \ifmmode ^{\rm m}
        \else $^{\rm m}$
        \fi
       }
\def\magd#1.#2{                          % magnitudes over decimal point
              \relax
              \ifmmode #1^{\rm m}
                       \hskip-0.55em.\hskip0.22em#2
              \else \hbox{#1$^{\rm m}
                    \hskip-0.55em.\hskip0.22em$#2}
              \fi
             }

\begin{document}

\title{Kinematics of ionized and molecular hydrogen in the core of M100}
\author{J. H. Knapen$^{1}$} 
\author{I. Shlosman$^2$} 
\author{C.H.~Heller$^{2,3}$} 
\author{R.J. Rand$^4$} 
\author{J.E. Beckman$^5$}
\and 
\author{M. Rozas$^5$}

\affil{$^1$ Department of Physical Sciences, University of Hertfordshire,
Hatfield, Herts AL10 9AB, UK.  E-mail knapen@star.herts.ac.uk}
\affil{$^2$ Department of Physics and Astronomy, University of Kentucky,
Lexington, KY 40506-0055, USA. E-mail shlosman,cheller@pa.uky.edu}
\affil{$^3$ Universit\"ats Sternwarte G\"ottingen, 11 Geismarlandstrasse,
37083 G\"ottingen, Germany}
\affil{$^4$ Department of Physics and Astronomy, University of New Mexico,
800 Yale Blvd NE, Albuquerque, NM 87131, USA}
\affil{$^5$ Instituto de Astrof\'\i sica de Canarias, E-38200 La Laguna,
Tenerife, Spain}

\journalid{Vol}{Journ. Date}
\articleid{start page}{end page}
\paperid{manuscript id}
\cpright{type}{year}
\ccc{code}
\lefthead{Knapen et al.}
\righthead{Kinematics in the core of M100}

\begin{abstract}

We present high angular and velocity resolution two-dimensional
kinematic observations in the spectral lines of \ha\ and CO $J=1
\rightarrow 0$ of the circumnuclear starburst region in the barred
spiral galaxy M100, and compare them with kinematics derived from our
previously published numerical modeling.  The \ha\ data,  fully
sampled and at sub-arcsecond resolution, show a rotation
curve that is rapidly rising in the central $\sim$140 pc, and stays
roughly constant, at the main disk value, further out. Non-circular
motions are  studied from the \ha\ and CO data by  detailed
consideration of the
velocity fields, residual velocity fields after subtraction of the
rotation curve, and sets of position-velocity diagrams. These motions
are interpreted as the kinematic signatures of gas streaming along the
inner part of the bar, and of density wave streaming motions across a
two-armed mini-spiral. Comparison with a two-dimensional velocity field
and rotation curve derived from our 1995 dynamical model shows good
qualitative and quantitative agreement for the circular and non-circular
kinematic components. Both morphology and kinematics of this region
require the presence of a double inner Lindblad Resonance in order to
explain the observed twisting of the near-infrared isophotes and the gas
velocity field. These are compatible with the presence of a global
density wave driven by the moderately strong stellar bar in this
galaxy. We review recent observational and modeling results on the
circumnuclear region in M100, and discuss the implications for bar
structure and gas dynamics in the core of M100 and other disk galaxies.

\end{abstract}

  %12.19.1

\vfill

Accepted for publication in the Astrophysical Journal, Volume 528 (Jan
1, 2000)

\section{Introduction}

The dynamics of circumnuclear regions (CNRs) is much more intricate than
the large-scale dynamics in disk galaxies. This complexity is due to a
variety of different factors, most notably (1), gas inflow and
modification of mass distribution towards the center due to
gravitational torques from bars and other non-axisymmetries; (2),
progressively increasing dynamical importance of gas within the central
few hundred pc (not only for star formation), compared with the rather
passive role it plays elsewhere; (3), formation of a multiple resonance
zone in the CNR which is sensitive to the varying underlying
gravitational potential; and, if appropriate, (4), nuclear activity
induced by the above dynamical processes.

Studies of the CNRs in disk galaxies are aimed at understanding the
causal relationship between galactic disks and their central activity,
both stellar and nonstellar. Observational and theoretical evidence
points towards the possibility that galaxies modify their mass
distribution on timescales short compared with the Hubble time (Friedli
\& Martinet 1993; Berentzen et al. 1998).  Redistribution occurs as a
result of intrinsic processes and also external interactions which lead
to gravitational torques acting on the gaseous and stellar disk
components (e.g., Combes 1988; Shlosman 1990; Zhang 1996).

Barred galaxies are examples of such dynamical and secular
evolution. They often experience starburst activity in their CNRs, with
star formation (SF) sites delineating nuclear rings within the central
kpc (see reviews by Kennicutt 1994; Buta \& Combes
1996). High-resolution imaging in the \ha\ emission line and at other
wavelengths usually brings out these structures consisting of molecular,
atomic and ionized gas and newly born stars, frequently subject to dust
extinction.

Cores of disk galaxies are places where resonances between stellar
orbital precession rates and pattern speeds of bars and ovals occur, at
about the rotational velocity turnover radius (e.g., Shlosman 1999 for a
recent review). The number of such inner Lindblad resonances (ILRs)
depends on the shape of the rotation curve, determined by the
axisymmetric part of the gravitational potential. If conditions for one
or two ILRs exist, they have a rather dramatic effect on the gas
response, i.e., forcing the gas streamlines to intersect and form pairs
of shocks. The radial gas inflow towards the center slows down, leading
to gas accumulation inside the outer ILR and to an increase in the SF
activity there (Schwarz 1984; Combes \& Gerin 1985; Shlosman, Frank \&
Begelman 1989; Knapen et al 1995a,b). In other words, the inner
resonances are precursors to the nuclear ring phenomenon.  The evolution
of the CNRs depends on a number of factors, such as the rate of gas
inflow along the stellar bar, the efficiency of SF in the nuclear rings,
and the dynamical importance of gas accumulating there.  Self-gravity in
the gas triggers SF in the nuclear rings and plays a major role in their
dynamical stability. More importantly, it also governs the rate of gas
``filtering'' across the resonance region (Shlosman, Begelman \& Frank
1990; Elmegreen 1994; Knapen et al. 1995b).  Observational testing of
theoretical predictions of gas kinematics in the CNRs of disk galaxies
is, therefore, of prime importance.

M100 has been classified by de Vaucouleurs et al. (1991; hereafter RC3)
as .SXS4. We have taken its distance as 16.1 Mpc (Ferrarese et
al. 1996), so that $1\sec$ corresponds to $\sim70$ parsec in the plane
of the galaxy.  This work focuses on observations and interpretations of
the two-dimensional CNR gas kinematics in M100 (=NGC~4321), a galaxy
with a bar of moderate strength. The observations are performed in the
spectral lines of \ha\ and CO $J = 1 \rightarrow 0$, using Fabry-P\'erot
and molecular interferometry techniques to achieve high angular and
velocity resolution. Our results are combined with high-resolution
optical and NIR $K$-band imaging and dynamical modeling of this CNR
(Knapen et al. 1995a,b; Shlosman 1996). The NIR core morphology in M100
(Knapen et al. 1995a; Ryder \& Knapen 1999; see also Fig.~8) is that of
a stellar bar encircled by a $7.5\sec-20\sec$ radius annular star
formation zone. The bar re-emerges at larger radii outside the ring. The
NIR isophotes are twisted in the CNR, by about 50\deg, in the leading
direction (to the bar) and back. A pair of short leading armlets is
visible at the ends of the observed inner bar structure.  The
circumnuclear SF occurs at the inner ILR (IILR), as indicated by the NIR
morphology and supported by our modeling (Knapen et al. 1995b). The
morphology in \ha\ (reproduced in Fig.~7) and blue light is strikingly
different from that in the NIR with the maxima of SF delineating a
tightly wound spiral structure, flanked by dust lanes, which connect
outward through the bar and to the main spiral arms in the disk. In $K$,
the outer incoming spiral arms are hardly discernible, and the inner
star-forming structure, though detectable (Ryder \& Knapen 1999), is
much less prominent.

The CNR of M100 was previously observed in detail using different
techniques. Broad-band optical morphological observations were first
presented by Morgan as long ago as 1958, and M100 figured in Sersic \&
Pastoriza's (1967) list of circumnuclear ``hot spot'' systems.  Similar
observations in \ha\ are given by Arsenault et al. (1988), Pogge (1989),
Cepa \& Beckman (1990), Knapen et al. (1995a) and Knapen (1998).

Previous kinematical studies include that of Arsenault et al. (1988) in
\ha, and interferometric observations in the CO $J=1 \rightarrow 0$
transition by Rand (1995), Sakamoto et al. (1995), and Garc\'\i
a-Burillo et al. (1998). Numerical modeling to account for the observed
structure and kinematics has been performed by Knapen et al.  (1995b),
Wada, Sakamoto \& Minezaki (1998), and Garc\'\i a-Burillo et al.
(1998). We shall deal comparatively with recent modeling, including our
own, and relate the observed kinematics presented here with the models
in Section 7.

M100 has received much observational and theoretical attention, not only
because it is one of the closest Virgo cluster galaxies, so its
structure can be well analyzed observationally, but also because its bar
of moderate strength gives rise to a particularly clear resonant
circumnuclear structure. Similar phenomenology is now recognized to be
fairly common in barred spirals, as reviewed by Buta \& Combes (1996),
and shown in papers by, e.g., Pogge (1990), Maoz et al. (1996),
Planesas, Colina \& P\'erez-Olea (1997), Elmegreen et al. (1997), Laine
et al. (1999), and P\'erez-Ram\'\i rez et al. (1999, in preparation).

This paper is structured as follows. New observations of the CNR in M100
are presented in Section~2, and the CNR morphology in \ha\ and CO is
discussed in Section~3. Circular and non-circular motions in the
observed region are analyzed in Sections~4 and 5, respectively.
Section~6 deals with the relevant results of numerical simulations and
Section~7 compares the simulations with observations, and our
interpretation with previously published studies. Conclusions are given
in the last Section.

\section{Observations}

\subsection{H$\alpha$}

We used the TAURUS~II instrument in Fabry-P\'erot (FP) mode on the
4.2m William Herschel Telescope on La Palma during the nights of
May 10 and 11, 1995. We windowed the TEK CCD camera to a size of
$600\times600$ pixels of 0.28 arcsec pixel$^{-1}$. The night was
photometric with $\sim0.7\sec$ seeing.  We observed two different
parts of the disk of M100, making sure that the central region of the
galaxy was present in both data sets.

The raw observations consist of a series of images at different
wavelengths, covering the \ha\ line emission from the galaxy. The
appropriately redshifted narrow band \ha\ filter ($\lambda_{\rm
c}=6601$\AA, $\Delta\lambda=15$\AA, using the galaxy's systemic velocity
$v_{\rm sys}=1586\kms$; RC3) was used as an order-sorting filter. We
performed wavelength and phase calibration by observing a calibration
lamp before and after each science exposure. We subtracted the
background sky value from each separate plane, and shifted the planes to
the same position using fits to foreground stars, before correcting the
raw data set using the {\sc taucal} software package. This produced two
data cubes of $600\times 600$ pixels $\times$ 55 ``planes" in
wavelength, separated by some 0.34 \AA, or 15.7 km\,s$^{-1}$. We adopted
a systemic velocity of $v_{\rm sys}=1586\kms$ following the RC3, even
though other determinations in the literature (e.g. Knapen et al. 1993)
give different values. Since this is a constant offset the choice of
$v_{\rm sys}$ will not have any influence on our results. Rozas et
al. (1998) used the same data set to determine velocity dispersions for
some 200 \hii\ regions across the face of the galaxy.

We combined the information on the central region of M100 from the two
individual data sets by cutting the region of overlap, of size
$200\times200$ pixels, from each cube, placing the sub-cubes at the same
grid position using fitted positions of foreground stars in the original
cubes, and of the nucleus of the galaxy, and averaging the individual
planes in the data cubes. The spatial resolution of the resulting data
set is $\sim$\secd 0.7$\!$. Astrometry was performed comparing the
positions of foreground stars and bright \hii\ region in the FP data
sets to their positions in the larger-scale \ha\ image of M100 from
Knapen (1998).

\begin{figure*}
\epsfxsize=15cm \epsfbox{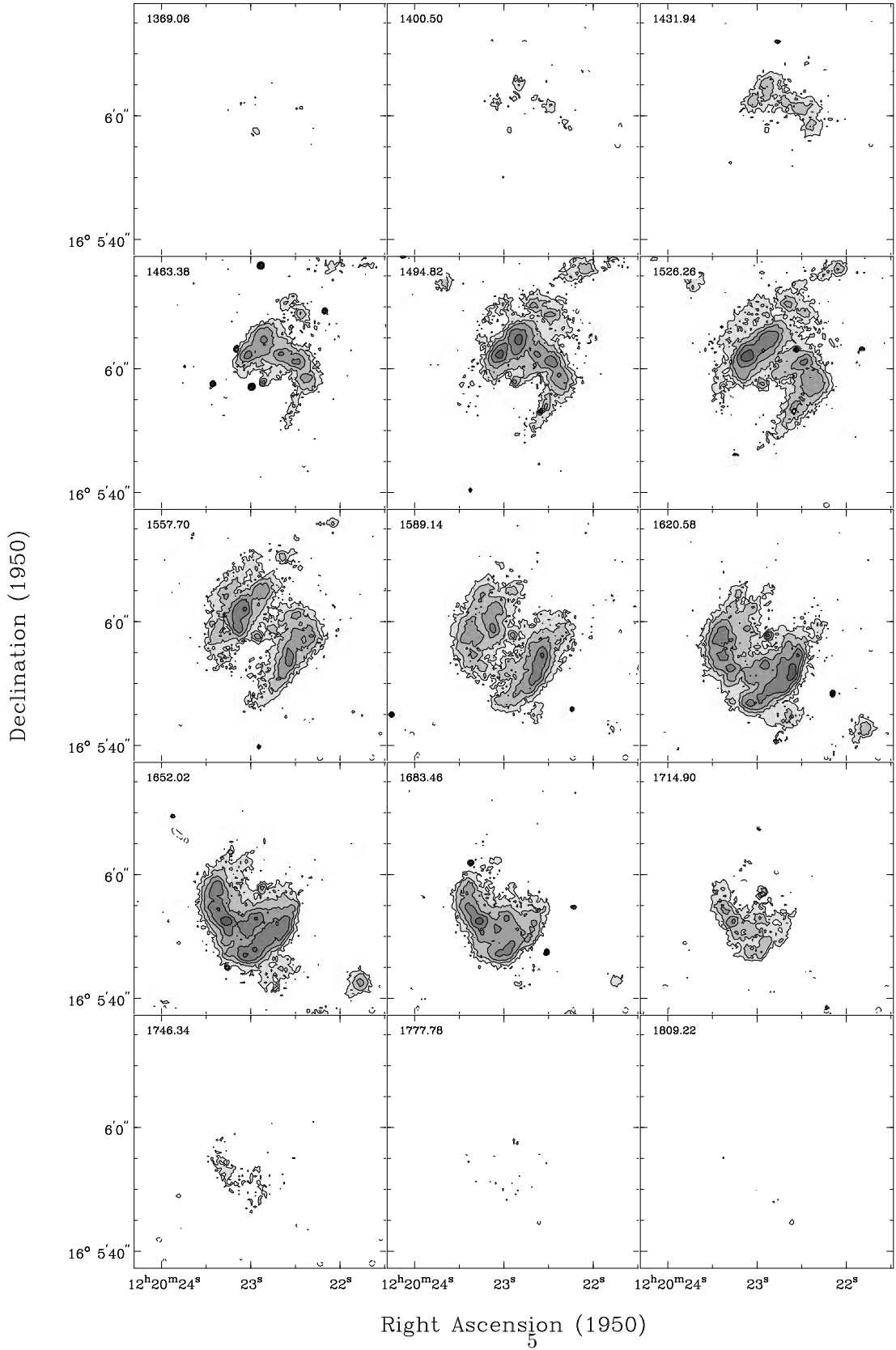}
\caption{Channel maps of the \ha\ emission from the CNR in
M100, at the full resolution, $\sim\secd0.7$.  Velocity of each channel
is indicated in the upper left corner.  Only every second channel is
shown.  Contour and gray levels are at approximately 4, 8, 16, 32 and 64
$\sigma$.}
\end{figure*}

We determined which channels of the data set were free of \ha\ line
emission after smoothing it to a resolution of $4\sec\times4\sec$.
Subsequently, we subtracted the continuum emission after fitting the
continuum to the 15 line-free channels on either side of the data
cube. The \ha\ emission as a function of increasing wavelength, or
velocity, is shown in a way equivalent to the standard channel maps in
radio astronomy in Fig.~1.  

\begin{figure*}
\epsfxsize=18cm \epsfbox{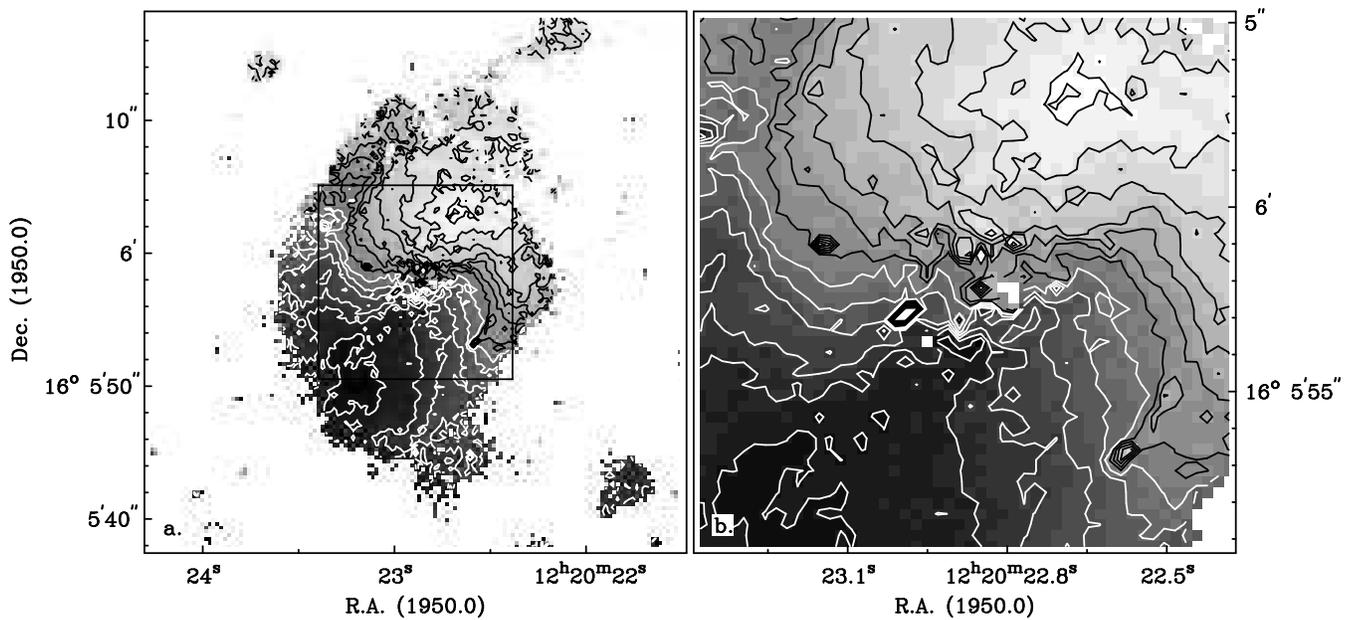}
\caption{H$\alpha$ FP velocity field (moment 1 map) of the central
region of M100, shown in contours and grey sales (left, {\it a.}). The
right panel ({\it b.}) is a close-up view of the central region. The
area covered by the right panel is indicated with a rectangle in the
left panel. Contours are separated by 15 \kms, from 1480\kms\ to
1675\kms. Black contours are low velocities, the first white contour is
at 1585\kms. The total intensity (moment 0) map is shown in Fig. 10a.}
\end{figure*}

 We have performed a moment analysis in the central region of M100, and
show the resulting velocity field (moment one map) in Fig.~2.  The
individual \ha\ profiles over all but the very innermost ($<1\sec$)
region are well suited for a moment analysis, double-peaked profiles
being practically absent. The detailed procedure used to produce the
moment maps is described in Knapen (1997).   The total intensity \ha\ map
(moment zero) as derived from the FP observations is shown in
Fig.~10a. It is comparable in quality to the narrow-band \ha\ image
published by Knapen et al. (1995a, see also Knapen 1998), with an
estimated spatial resolution of \secd 0.6 -- \secd 0.7$\!$.

In order to check aspects of the distribution and kinematics of the
ionized gas directly with those of the molecular gas, we produced an
\ha\ data set with the spatial resolution of the CO data ($\secd
4.6\times\secd3.6$) by convolving the original data cube with the
appropriate Gaussian. Subsequent analysis of this smoothed data set was
in all cases completely analogous to that of the full resolution data
set.

\subsection{CO}

\begin{figure*}
\epsfxsize=17cm \epsfbox{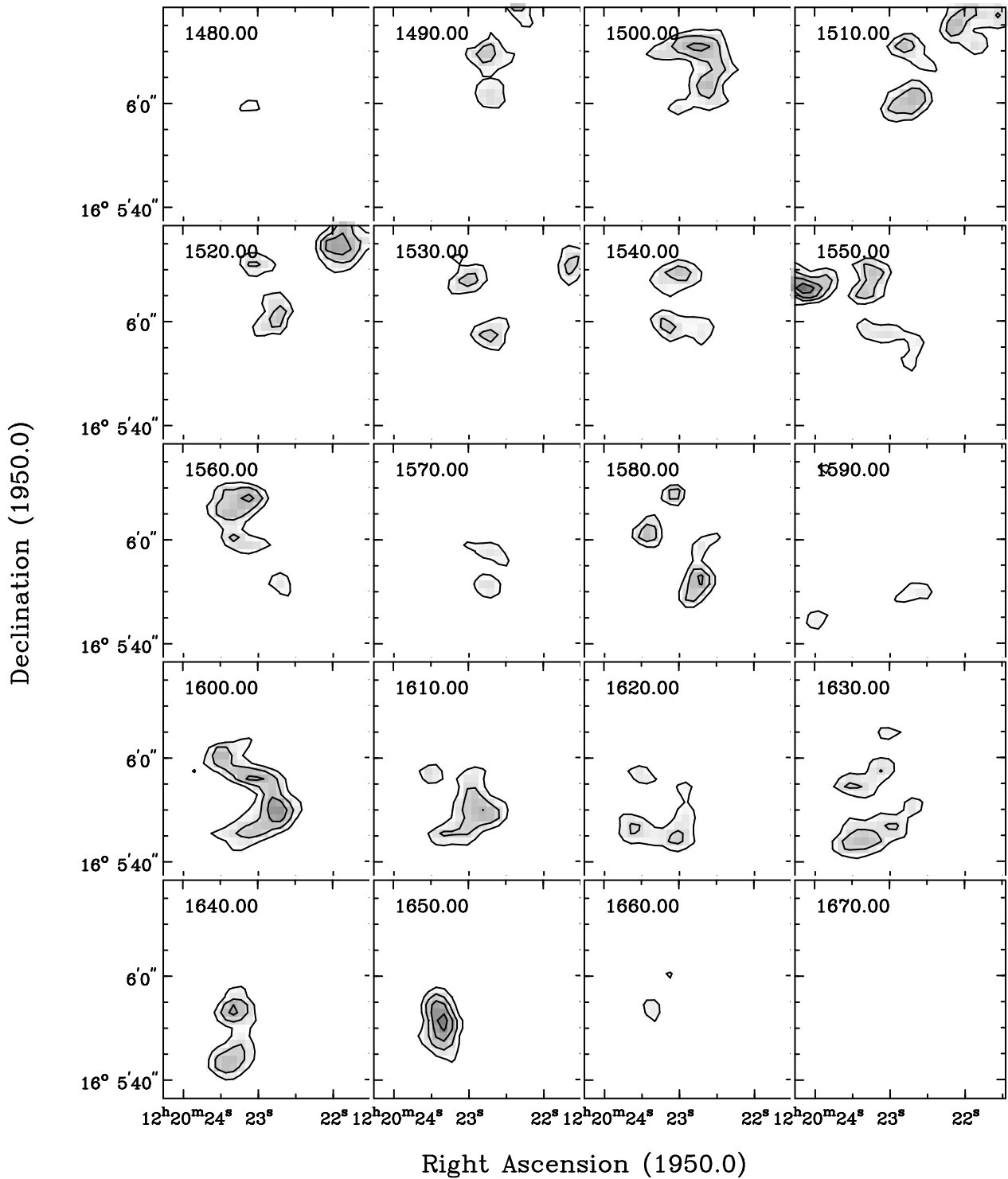}
\caption{Channel maps of the CO emission from the CNR in
M100, at the same scale and orientation as those of \ha\ in Fig.~1.
Velocity of each channel is indicated in the upper left
corner. Contour levels are at 0.6, 1.1, 1.7 and 2.2 K}
\end{figure*}

\begin{figure}
\epsfxsize=8cm \epsfbox{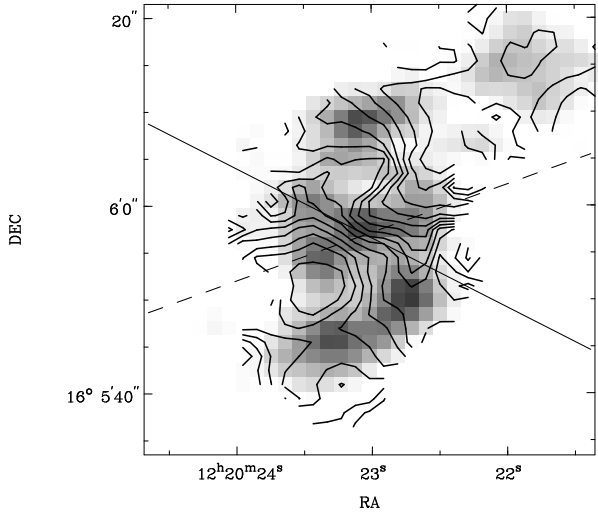}
\caption{Contours of the CO velocity field of the 
central region of M100, overlaid on a grey-scale representation of the
total intensity CO map. The drawn line indicates the kinematic minor
axis, while the dashed line is the position angle of the large-scale
bar. Contours are from 1500\kms\ to 1680\kms, in steps of
10\kms. Epoch is B1950.0.}
\end{figure}

The CO observations used are those described by Rand (1995), who
observed the central region of M100 with the BIMA\footnote{Operated
by the University of California, Berkeley, the University of Illinois,
and the University of Maryland, with support from the National Science
Foundation} interferometer in the CO $J=1 \rightarrow 0$
transition. Channel maps of the CO data in the CNR of M100 are shown
in Fig.~3. We re-analyzed some of these data, concentrating on the gas
kinematics in the CNR. We first removed noise peaks outside the area
where CO emission is expected by setting pixel values at those
positions to a ``blank'' value. This was done interactively by inspecting the
individual channel maps, continually comparing with the same and
adjacent channels in a smoothed data cube, in a way similar to the
procedure described by Knapen (1997).  This new data set was used
to calculate the total intensity and velocity moment maps. The
resulting moment images show the same structure as published by Rand
(1995) but are somewhat more sensitive in the central region. We show
the velocity field thus produced, overlaid on a grey-scale
representation of the total CO intensity, in Fig.~4. The spatial
resolution in the CO data set is $\secd 4.6 \times \secd 3.6$ (PA
88\deg), the channel width is 5 \kms, and the channel noise is about
40 mJy\,beam$^{-1}$.

We detect 244 Jy\,km\,s$^{-1}$ ($\pm10\%$), which is
40\% of the flux measured by Kenney \& Young (1988) within the central
45\sec, of $626 \pm 119$ Jy\,km\,s$^{-1}$. Using a standard conversion
factor, which may well not hold in this environment (e.g. Rand 1995),
our flux would correspond to a mass of about $10^9$ M$_\odot$, and gas
surface densities for four individual clumps in the CNR to $\sim
220\pm20$\,M$_\odot$\,pc$^{-2}$. Presumably due to a different
resolution and/or calibration, Sakamoto et al. (1995) find somewhat
higher densities, which they estimate to exceed the critical density for
nuclear ring fragmentation (the Toomre criterion). However,
uncertainties not only in the CO to H$_2$ conversion factor, but also in
the estimates for the epicyclic frequency and the dispersion velocity of
molecular clouds in the ring, most likely far outweigh the difference
between the observed gas surface density and the estimated critical
density, and we therefore do not attempt a similar analysis here.

\section{Core Morphology}

\subsection{\ha}

The \ha\ morphology of the CNR in M100 is that of an incomplete ring
extending from 7.5\sec\ to 20\sec\ from the center; it has been
described in detail by Knapen et al. (1995a,b). We summarize only the
most important features here, which can all be recognized in the total
intensity \ha\ image (Fig.~10a). The nuclear ring emits about 16\% of
the total \ha\ emission from the galaxy. Knapen (1998) catalogues almost
2,000 separate \hii\ regions across the galaxy, of which only 99 are
found in the CNR, predominantly bright ones.  Crowding is a problem
which leads to an underestimate of the number of, especially less
luminous, \hii\ regions in the CNR, but it is clear that a relatively
large amount of \ha\ emission is produced in a small volume, directly
indicating a high rate of massive SF. The SF in the nuclear ring is
organized into four complexes corresponding to the maxima of the gas
compression at the ``twin peaks'' at the bar's minor axis, and at the
ends of a pair of small spiral armlets at the bar's major axis. These
four complexes of SF were identified by Knapen et al.  (1995a; their
fig.~1c), and shown to be heavily biased towards the position of the
IILR, where the gas is prone to local Jeans instabilities (Knapen et
al. 1995b).

\subsection{Molecular gas}

\begin{figure}
\centerline{\psfig{figure={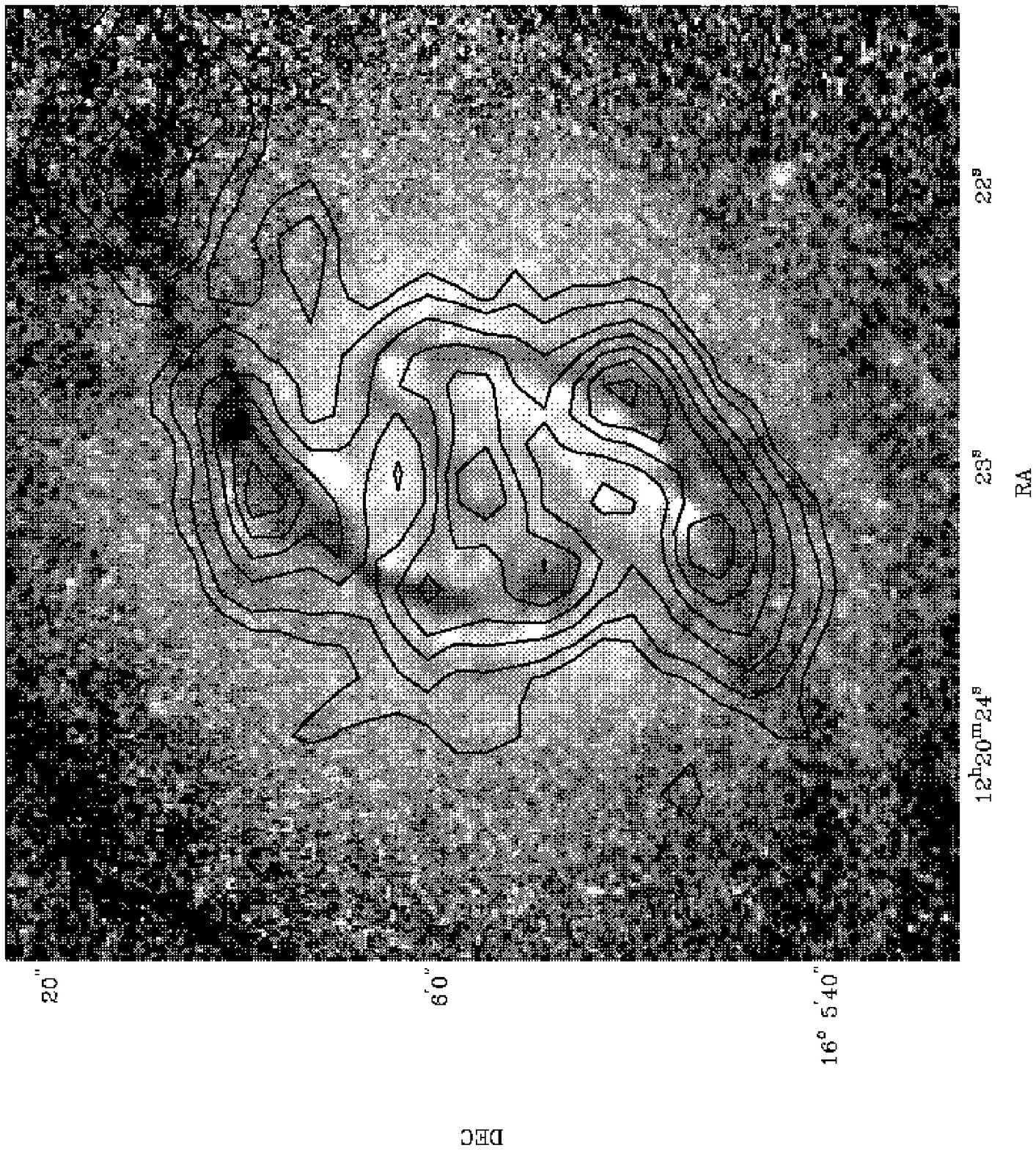},width=8cm,angle=-90}}
\caption{Contour representation of the CO total intensity map of Fig.~3,
overlaid on the $I-K$ color index map of the CNR in M100
from Knapen et al. (1995a).
Contours are 0.25, 0.5, 1, 1.5, 2, 2.5, and 3 $\times 27.8$
K\,km\,s$^{-1}$. Nucleus is indicated with a cross. Local minima in CO can
be recognized by comparison with the grayscale image in Fig.~4. As in
Fig.~4, epoch is B1950.0.}
\end{figure}

\begin{figure}
\centerline{\psfig{figure={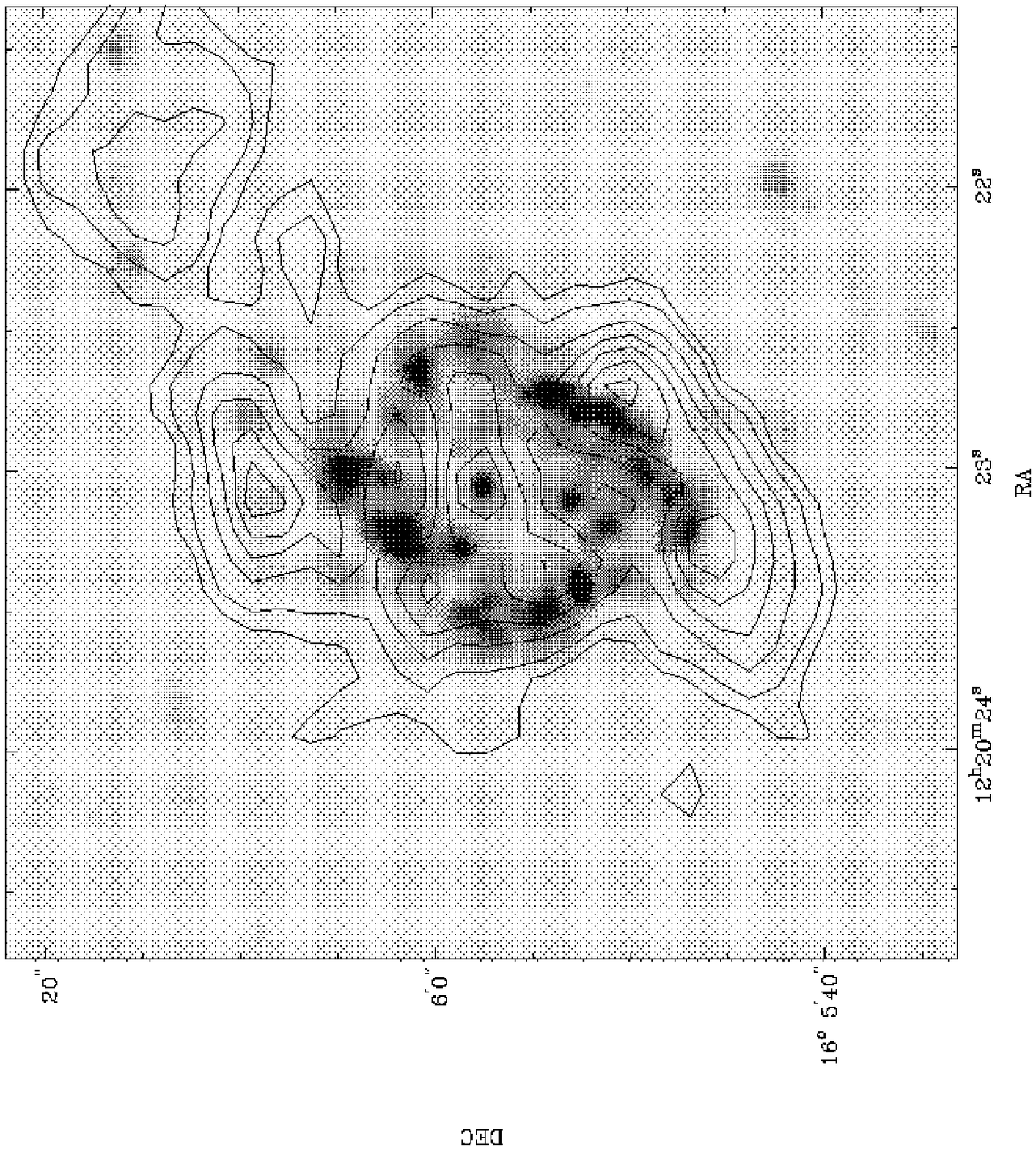},width=8cm,angle=-90}}
\caption{As Fig.~5, now with CO contours overlaid on the  \ha\ image of 
Knapen et al. (1995a).}
\end{figure}

\begin{figure}
\centerline{\psfig{figure={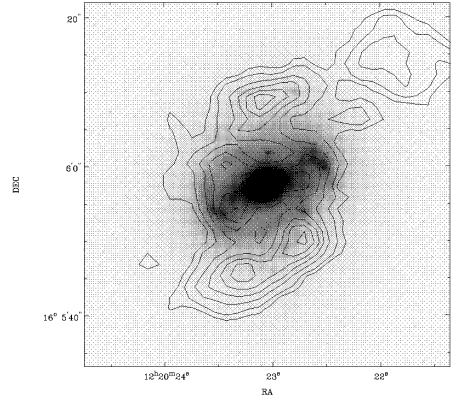},width=8cm,angle=-90}}
\caption{As Fig.~5, now with CO overlaid on the $K$ image of Knapen et al.
(1995a).} 
\end{figure}

The distribution of molecular gas in the CNR of M100 has been described
by Rand (1995) and by Sakamoto et al. (1995), while Knapen et al.
(1995b) showed older CO data of Canzian (1990), at lower resolution, and
compared its distribution with optical-NIR tracers. In Fig.~4 we present
the CO total intensity and velocity distributions as obtained from the
re-analyzed Rand (1995) data, and in Figs.~5, 6 and 7 the CO
distribution is overlaid on $I-K$ (from Knapen et al. 1995a), \ha\ (as
in Fig.~10a) and $K$ (Knapen et al. 1995a) images. The basic notions are
clear from this collection of images: the CO is distributed along two
spiral arms, coinciding with the dust lanes along the SF spiral arms as
seen in \ha. The CO along the arms is clumpy, with a local maximum of CO
emission centered on the nucleus. The CO spiral arms follow the dust
lanes out into the large-scale bar, and into the disk of the galaxy
(Rand 1995).  Our new map confirms observations and numerical
simulations by Knapen et al. (1995a,b) that there are local CO peaks
located slightly upstream from the four main maxima of SF (upstream in
this context means counterclockwise, assuming that the disk and CNR
spiral arms are trailing). This is also supported by Sakamoto et
al. (1995). 

The pair of leading arms, observed in $K$ inside the star-forming
ring-like region by Knapen et al. 1995a, is visible in the CO
morphology. These leading arms are also visible in the CO maps published
by Rand (1995) and Garc\'\i a-Burillo et al. (1998), in both cases more
clearly on the west than on the east side of the nucleus. As explained
in Knapen et al. (1995a,b) such leading arms are signatures of a double
ILR and are not expected under any different circumstances.  The
$K$-band image of Ryder \& Knapen (1999), of much better angular
resolution than the original $K$ image in Knapen et al. (1995a),
confirms the existence of the leading armlets. The suggestion by
Garc\'\i a-Burillo et al. (1998) that their appearance is at least
partly a result of dust extinction effects is not supported, because
these authors publish nor their $A_V$ map, nor the images resulting from
the exinction correction based upon it. Knapen et al. (1995a) explained
in detail the procedure used to correct the $K$-image for extinction by
cold dust, and the underlying assumptions. This exercise resulted in a
clearer rather than less obvious view of the leading arms, as well as of
the peaks of SF, K1 and K2 (fig.~1d of Knapen et al. 1995a). The leading
arms are part of the general twisting of NIR isophotes by an angle
of $50^\circ$ in the leading direction (to the bar) and back. 

\section{Kinematic results: circular motions}

\subsection{\ha}

The velocity field of the central region (Fig.~2) shows the basic
``spider'' pattern indicating circular motions in a rotating disk, with
rapidly increasing rotation velocities, but also important deviations
from circular velocities.  The first effect to note specifically is the
{\sf S}-shaped deviation of the iso-velocity contours near the minor
axis, indicating gas streaming along the inner part of the bar. Another
important deviation in the velocity field is due to a bar-induced
spiral density wave, and seen most clearly toward the NE and SW of the
nucleus, at radii of some 9\sec. These deviations from circular motions
are discussed in detail below (Sect.~5).

In order to describe the circular motions we have derived rotation
curves from the \ha\ velocity fields at both full ($\sim$\secd 0.7) and
low (as in CO) resolutions. The exact procedure used is based on the one
described by Begeman (1989; see also Knapen 1997). In short, parameters
describing the galaxy's velocity field in terms of a set of concentric
rings are fitted to the data using a least-squares algorithm. In the
fit, we fixed the position of the center, and the inclination angle $i$
(at $i=30\deg$). The fits are robust against changes in the central
position, but fixing the central position after having determined it in
preliminary fits increased the accuracy of the rotation curve fit. The
astrometry of the \ha\ data set is not good enough to give an accurate
kinematic central position, however, the kinematic and morphological
center positions as determined from the \ha\ data agree to within the
errors. We adopted a value for the inclination of the inner disk very
close to the one used in our previous \hi\ study of the main disk of the
galaxy (Knapen et al. 1993), in the absence of evidence to the
contrary. Data points within 30\deg\ of the minor axis were excluded
from the fits.

\begin{figure}
\epsfxsize=8cm \epsfbox{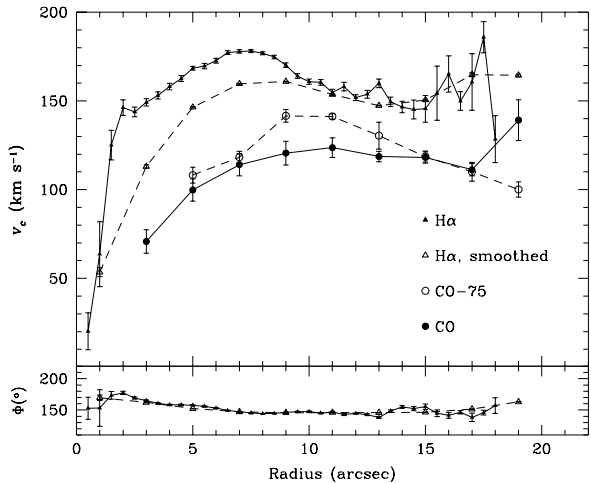}
\caption{Rotation curve derived from the \ha\ and CO data sets for the
inner region in M100 (top panel), and the fitted run of the position
angle of the major axis ($\Phi$; \ha\ only) against radius. The various
curves are \ha\ full resolution (filled triangles), \ha\ smoothed to the
CO resolution (open triangles), and CO (filled dots). Also included
(open dots, dashed line) is the CO curve obtained when only points in
the velocity field at more than 75\deg\ from the minor axis are taken
into account for the rotation curve fit.}
\end{figure}

The rotation curve was fitted at \secd 0.5 intervals in radius from the
center to a radius of 18\sec. The fitted values for the major axis
position angle $\Phi$ and the circular velocity $v_{\rm c}$ are
presented as a function of radius in Fig.~8. The fitted value for the
systemic velocity of the galaxy reproduces, within the errors of the
fit, the value introduced during the calibration of the data. Using
exactly the same parameters we also fitted a rotation curve to the
smoothed data set, at $\secd 4.6 \times \secd 3.6$ resolution. This
curve, and the radial run of $\Phi$, are also shown in Fig.~8.

The rotation curve at full resolution shows a very steep rise (to almost
$v_{\rm c}=150\kms$ in 2\sec, or $\sim140$ pc), which must in fact be a
lower limit to the true rise of the rotation velocities due to the
fitting algorithm used (see, e.g., Sofue et al. 1997). We checked whether
this steep rise could be due to the instrumental effect of less than
perfect continuum subtraction in the nuclear region (which shows up to
some degree in the original data) by setting the data values in the
central $\sim1\sec$ and $2\sec$ radius to ``blank'' in the
continuum-subtracted data set, and producing new versions of the first
moment (velocity) map and rotation curve fit. In both
cases the rise of the rotation curve did not get shallower (it steepened
within the errors of the fit in the 2\sec\ case) and, therefore, the
steep rise observed is not due to instrumental effects. It corresponds
to the equivalent of a spherically distributed mass of $\sim 10^9\ {\rm
M_\odot}$ within the inner 140 pc (in agreement with results obtained by
Sakamoto et al. 1995).

The observed \ha\ rotation curve reaches a local maximum at around
$v_{\rm c}=180\kms$ at a radius of $\sim7\sec$, the radial range where
the bulk of the \ha\ emission from the spiral armlets is found.  The
drop in $v_{\rm c}$ at larger radii occurs where the density wave
streaming motions are strongest (Fig.~2, see also Sect.~5) and is in
fact a result of the noncircular motions.  The position angle of the
major axis is $\Phi=147\deg\pm3\deg$ for radii over 6\sec, just a few
degrees different from the disk value of $153\deg\pm1\deg$ derived from
\hi\ aperture synthesis observations (Knapen et al.  1993). 

The rotation curve derived from the $\secd 4.6\times \secd 3.6$
resolution velocity field confirms the above findings.  Differences are
due to the lower angular resolution: the initial rise is slower
and the peak at $R=7$\sec\ has slightly shifted.

In Fig.~10d we show a position-velocity (LV) diagram of the
full-resolution \ha\ data along the major axis. A series of six
different slices were produced from the cube, offset by half a
resolution element (\secd 0.4) and parallel to the major axis. These
six slices were then median-averaged to produce the LV-diagram
shown. The diagram shows the original data, without any profile or
ring fitting such as done in the process of producing the rotation
curve, yet it confirms the main findings from our rotation curve
analysis, as described above.

\subsection{CO}

We also derived a rotation curve from the CO velocity field, using the
same procedure as for the \ha. However, due to the different
characteristics of the data set, we excluded data points within
20\deg\ from the minor axis, fixed the position angle of the major
axis as $\Phi=153\deg$, and calculated rotation curve data points with
\secd 1.5 spacing in radius. The resulting rotation curve is shown in
Fig.~8, compared to the \ha\ rotation curve at similar spatial
resolution to the CO, as derived in the previous section.

\begin{figure*}
{\centering \leavevmode \epsfxsize=.45\textwidth \epsfbox{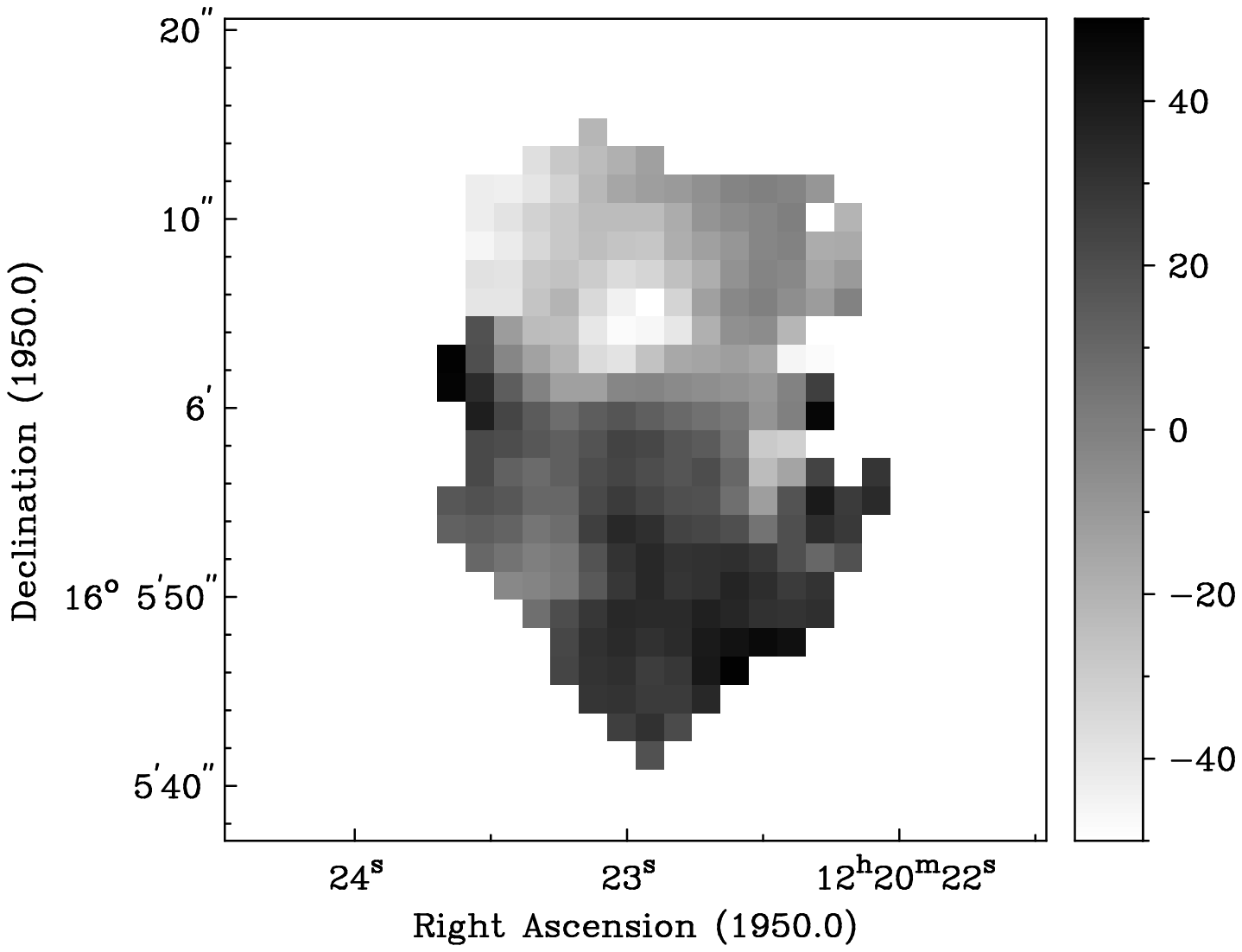} 
\hfil \epsfxsize=.45\textwidth \epsfbox{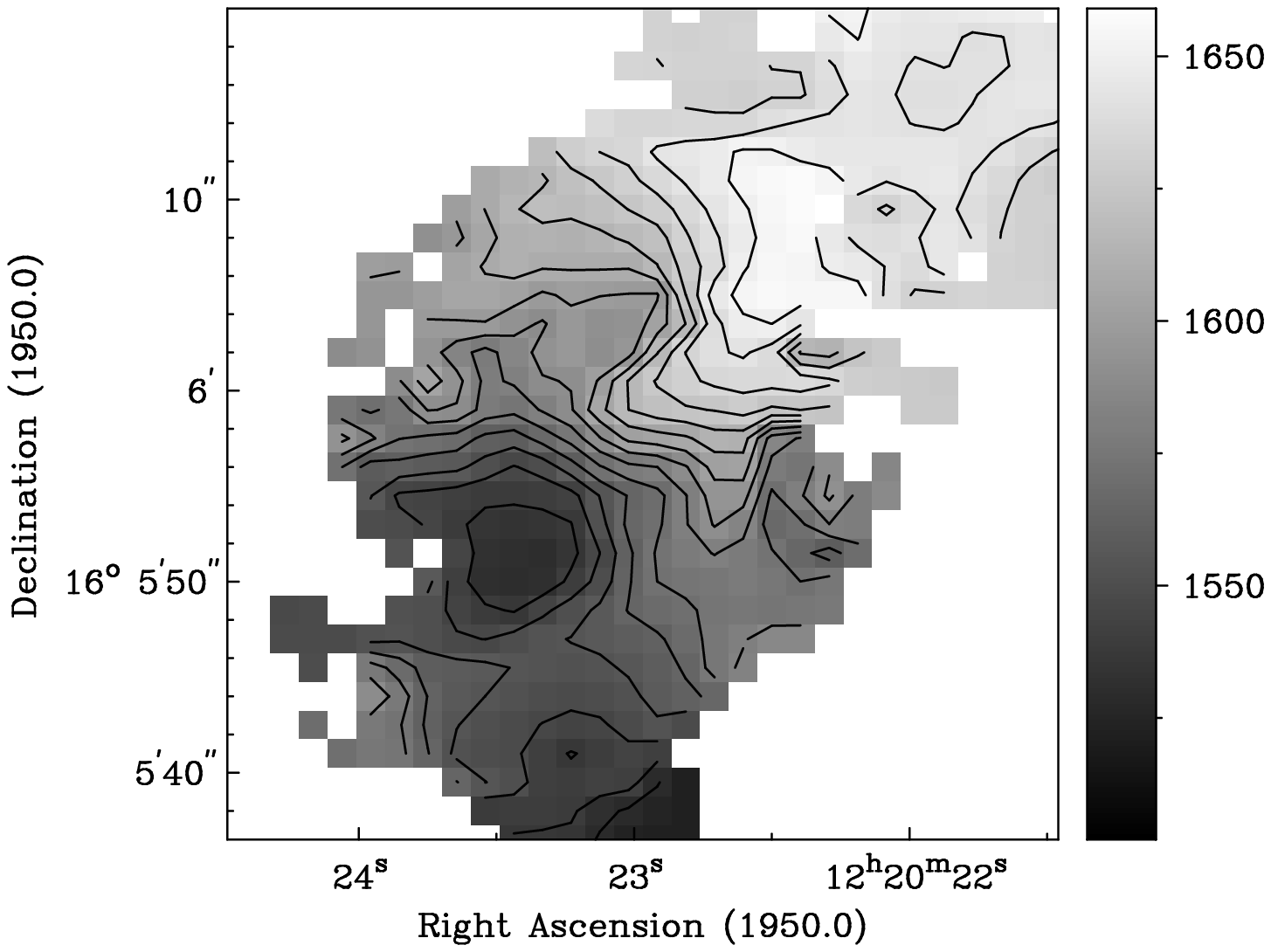}}
\caption{{\it a.} (left) Difference map of the (smoothed) \ha\ minus the
CO velocity fields.  The wedge on the right hand side indicates how
differences in \kms\ correspond to grey scales, positive differences
indicate relatively more velocity in \ha.  {\it b.} (right) CO velocity
field, shown for comparison, and at the same scale as part {\it a}. 
Grey scales are mapped to velocities in \kms\ by the wedge on the right,
and contour levels are 1500 to 1680 \kms\ in steps of 10 \kms, like in
Fig.~4.}
\end{figure*}

There are two very notable differences between the CO and H$\alpha$
rotation curves: the CO curve rises much more slowly than the
H$\alpha$ curve and reaches much lower maximum ``circular''
velocities.  These differences are largely due to spiral arm streaming
motions (discussed in Sect. 5) in combination with the offset of
the CO and H$\alpha$ inner spiral arms.  In particular, examination of
a difference map of the two velocity fields (Fig.~9, where this map is
shown along with the CO velocity field at the same scale) shows that the
main region of discrepancy is close to the beginning of these two arms
where they join onto the inner bar (i.e., inside the nuclear ring),
within about 60 degrees of the minor axis.  Two major SF regions are
found here. In these regions the CO and H$\alpha$ velocities differ by
about $40-80 \kms$ in the plane of the galaxy. Since both CO and \ha\
measure basically gas dynamics (\ha\ because the young massive stars
that cause the ionization of the hydrogen are still on the gas
orbits of their parent molecular clouds), the observed differences are
most probably due to the different spatial distributions of the emission.

To demonstrate this effect, we compare rather extreme rotation curves,
formed by excluding points less than 75$\deg$ from the minor axis. The
result for CO is shown in Fig.~8; the result for \ha\  is very similar
to the whole-disk rotation curve in Fig.~8.  With the combined
influence of the streaming motions and spatial offsets minimized, the
rotation curves show much better, although still not perfect,
agreement.

This effect of streaming motions on rotation curves was also noted for
M51 by Rand (1993).  It highlights the difficulties in deriving a
correct axisymmetric rotation curve when strong streaming is present
and emission is biased to a small range of spiral phase. As discussed
further below, these difficulties have important implications for the
use of rotation curves in the derivation of galactic potentials for
subsequent modeling. 

\section{Kinematic results: deviations from circular motions}

In order to demonstrate the nature of the streaming motions observed in
CO and \ha, we use three different techniques. Firstly, (Sect.~5.1) a
mostly qualitative description of the velocity fields, secondly
(Sect.~5.2) a study of residual velocity fields obtained after
subtracting a two dimensional version of the rotation curve from the
observed velocity field (most useful for the \ha\ data), and thirdly
(Sect.~5.3) an analysis of series of position-velocity diagrams along
and parallel to the inner bar minor axis (most useful for the CO data).

\subsection{Velocity fields}

From the velocity maps in \ha\ and CO (Figs.~2, 4) one can easily
recognize deviations from circular motions due to two distinct
dynamical processes: spiral density wave streaming motions 
associated with the armlets and gas motions along the inner part of 
the bar at smaller radii. 

Density wave streaming motions are recognizable in the deviations from
the regular shape of the velocity contours (isovels), especially toward
the NE and SW of the nucleus near radii of some $9\sec-10\sec$, and most
clearly in \ha\ due to the higher resolution. They are strongest where
we previously inferred the position of the incoming spiral arms, just
outside the well-defined dust lanes (Knapen et al. 1995a,b; this is also
just outside the radius of the strong \ha-emitting regions lying along
the inner bar's minor axis, as can be best seen in  a comparison of
Fig.~2 and Fig.~10a). Although
the signature of the streaming motions is most obviously visible near
the minor axis of the galaxy, they can in fact be recognized
consistently out to some 60\deg\ on either side of the minor axis. As
estimated from the \ha\ velocity contours, the (projected) excess or
streaming velocities are of the order of 40\kms. This kinematic
detection confirms that indeed the incoming spiral arms are part of the
grand-design density wave system and not a collection of flocculent
spiral arm fragments (as proposed by Pogge 1989 and Cepa \& Beckman
1990).

In the central $\sim5$\sec, again seen most clearly in \ha\ but also in
CO, the isovels do not run parallel to the minor axis as would be the
case for purely circular motions, but show a deviation characteristic of
gas streaming along the preferred major axis of the bar, as expected
(Roberts, Huntley  \& van Albada 1979) and observed (e.g. Bosma 1981) in
barred galaxies. The deviations seen here in \ha\ and CO occur on the
scale of the inner bar-like feature as seen in the NIR, and are 
confirmed by the numerical modeling (see Sect~6).

The deviations from circular motion due to both the spiral arms and
inside the nuclear ring have been found in CO and briefly described by
Rand (1995) and Sakamoto et al. (1995).  The spiral arm streaming
motions can also be seen in the CO map of Garc\'\i a-Burillo et
al. (1998), and are mentioned briefly by these authors. Arsenault et
al. (1988) noted these streaming motions in their \ha\ Fabry-P\'erot
velocity field, although the spatial resolution of their data was not
adequate to observe any details.

\subsection{Residual velocities}

\begin{figure*}
\epsfxsize=17cm \epsfbox{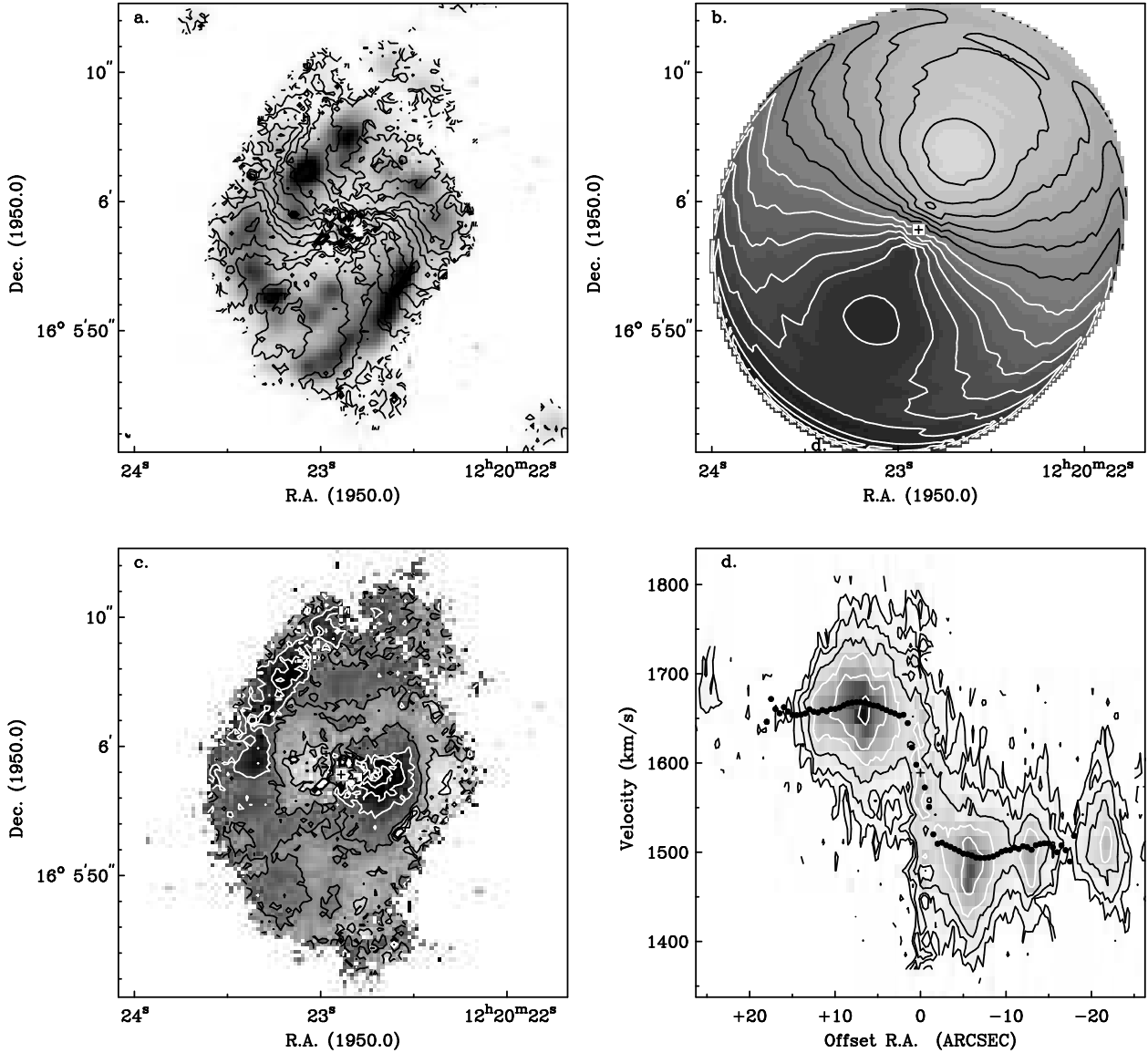}
\caption{{\it a.} (upper left) \ha\ velocity field of the circumnuclear
region of M100 at full ($\sim \secd 0.7$) resolution, overlaid on a
grayscale representation of the \ha\ total intensity, or moment zero,
map of the same region.  Contour levels are as in Fig.~2. 
{\it b.} (upper right) Model velocity field as determined from a
rotation curve with a constant position angle of 153\deg\ (see text). 
Contour and gray levels as in Fig.~10a. {\it
c.} (lower left) Residual velocity map, obtained by subtracting the
model (Fig.~10b) from the velocity field (Fig.~10a). Contours are at
$-45, -30$ and $-15$\,\kms\ (black) and 0, 15, 30 and 45\,\kms\
(white), with grayscales indicating a slightly smaller range from $-30$
to 30\,\kms, and higher values
coded darker. {\it d.}  (lower right) Position-velocity diagram along
the major axis ($\phi=153\deg$) of the \ha\ FP data set. Contour levels
are at approximately $-2\sigma$ (dashed), 2, 4, 8 (black), 16, 32, and
64$\sigma$ (white). Overlaid (black dots) is the rotation
curve for the whole disk at the same resolution. The cross marks the
kinematic center.}
\end{figure*}

One of the most direct ways to demonstrate the position and amplitude
of deviations from circular motion is by converting the derived
rotation curve into a two-dimensional axisymmetric velocity model
(shown for \ha\ in Fig.~10b), and subtracting it from the observed
velocity field (Fig.~10a). The residual \ha\ velocity field thus
obtained (Fig.~10c) shows qualitatively where non-circular motions
occur, and where they are strongest. A problem with this approach is
that each point on the rotation curve is produced by azimuthally
averaging the complete velocity information within a certain radial
range. A certain point or range of points on the rotation curve can
thus be directly influenced by non-circular motions, such as the ones
discussed above, preferentially occurring within the specific radial
band or bands corresponding to that point.  The rotation curve, and
consequently the model velocity field, will thus contain contributions
from non-circular motions, which will therefore not appear to their
full extent in the residual map. Only position-velocity diagrams
(Sect.~5.3) can reproduce the full available information, but since
residual velocity fields are so much easier to interpret, we present
and discuss those first. In order to minimize the impact of the
problem described above, we have used a slightly modified version of
the rotation curve for this part of our analysis. Instead of leaving
the position angle of the major axis a free parameter in the rotation
curve fitting (as in Sect.~4) we now fix the position angle at a value
of 153\deg. This results in a model velocity field (Fig.~10b) that is
less influenced by non-circular motions.

The residual \ha\ velocity map ( Fig.~10c) confirms that the most
important regions of excess residual velocity are related to the
innermost bar, and the incoming spiral arms. The E and W sides of the
inner bar show positive (centered at RA \hmsd 12h20m23.12s , dec \dms
16d06m1s in Fig.~10) and negative (\hmsd 12h20m22.60s , \dms
16d05m58s) residual velocities, respectively, with the spiral arms to
the NE and SW exhibiting similar behavior (positive residual velocity
component centered at RA \hmsd 12h20m22.43s , dec \dms 16d05m4s;
negative at \hmsd 12h20m23.15s , \dms 16d06m6s). This is expected for
a twofold symmetric deviation from the circular velocities.
The magnitude of the residual velocities is a few
tens of \kms, lower than the values obtained from the isovelocity
contours in e.g. Fig.~10a or the analysis in Sect.~5.3, as expected
(see previous paragraph).

This type of behavior along the major axis of the residual velocity
field is interesting: at first slightly negative, then slightly positive
when going to larger radii, to both the NW and SE. It indicates that,
presumably due to the non-circular motions, the rotation curve as
derived from the whole velocity field at first slightly overestimates
the circular velocity, then slightly underestimates it. The effect in
\ha\ is small, but is exactly the same as highlighted in CO when we
derived a rotation curve from a very limited part of the velocity field
close to the major axis (Fig.~8; Sect.~4.2), indicating the validity of
the approach followed in Sect.~4.2, and of the conclusions drawn there
about the cause for the differences between the \ha\ and CO rotation
curves.

\subsection{Position-velocity diagrams}

\begin{figure*}
\epsfxsize=15cm \epsfbox{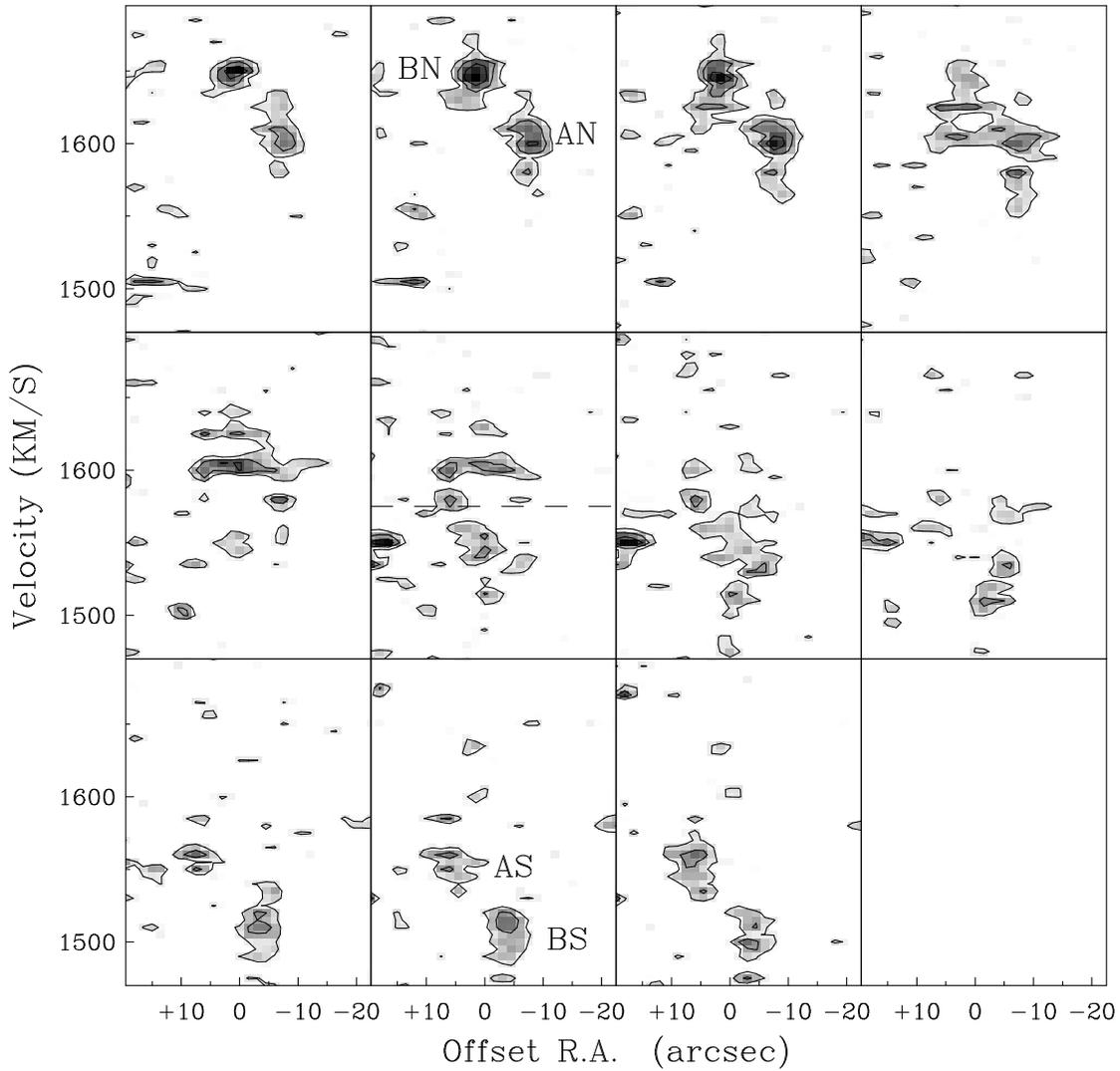}
\caption{Position-velocity diagrams in CO along (sixth panel, indicated
by a horizontal dashed line) and
parallel to the minor axis in the central region of M100, position
angle 63\deg. Panels to the left of and above the minor axis panel are
cuts north of the minor axis, panels to the right of and below are
south. Cuts are separated by \secd 1.5, or about half a beam. The
systemic velocity of the galaxy is 1575 \kms, indicated by the
horizontal line in the minor axis panel. Named features are
discussed in the text. Contour levels are 0.63, 1.25 and 1.88 K, gray 
levels span the same range.}
\end{figure*}

\begin{figure*}
\centerline{\psfig{figure={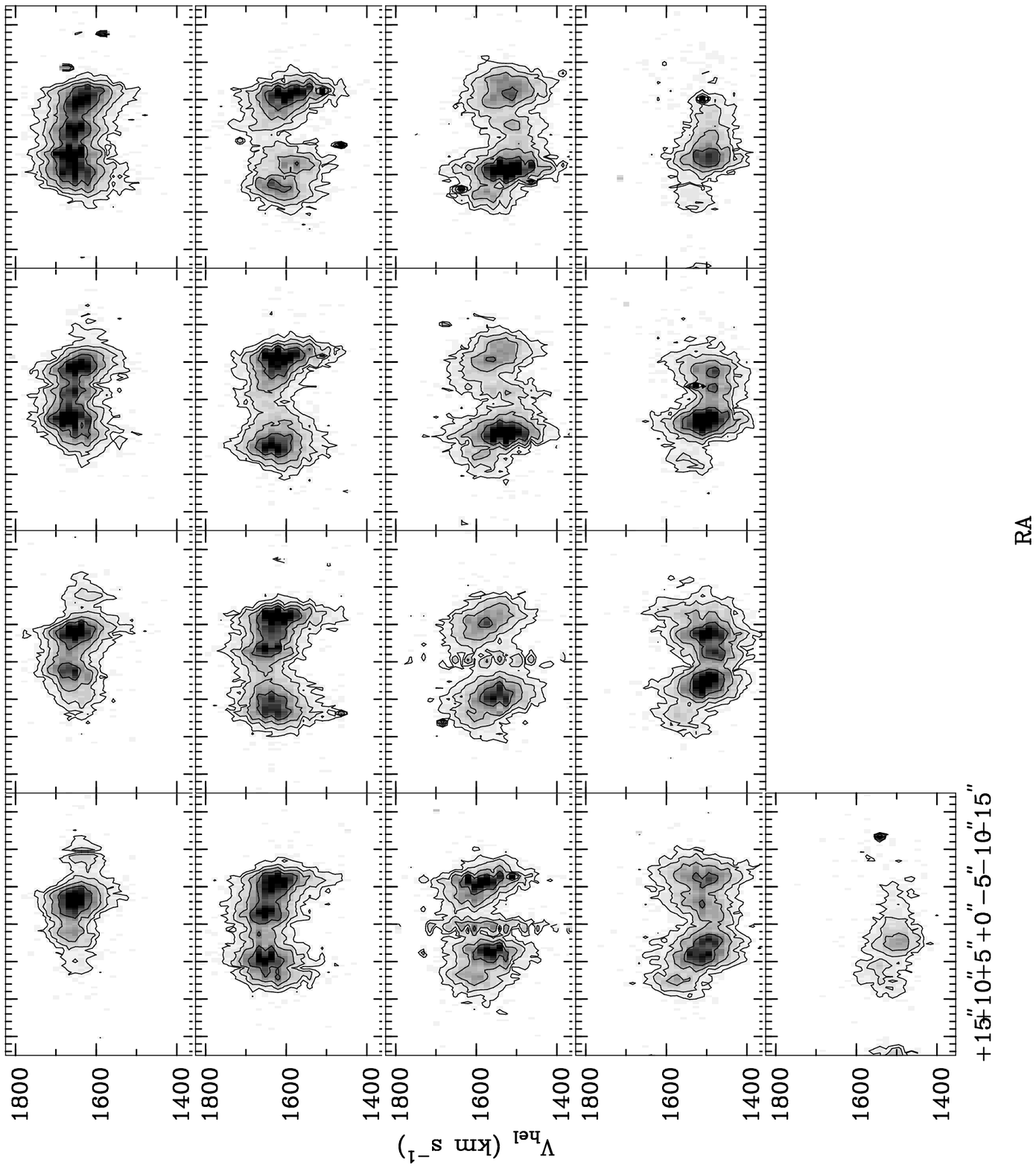},width=\textwidth,angle=-90}}
\caption{As Fig.~11, now for the \ha\ data at full resolution. The 
separation between the individual position-velocity diagrams is \secd
0.4 (about half a beam), but only every fourth diagram is shown. The
position-velocity diagram along the minor axis is the middle one in
the left row. Contour levels are at 3, 6, 12 and 24$\sigma$, the
highest (black) gray level is reached at 43$\sigma$.}
\end{figure*}

It is generally dangerous to use moment maps for interpretation of
kinematical features in regions where profiles may deviate from a
Gaussian shape, and/or have multiple peaks. Since the inner region under
consideration here is clearly such a region of enhanced risk, we
produced a set of position-velocity diagrams along and parallel to the
minor axis, shown in Fig.~11 (CO) and 12 (\ha).  It is immediately clear
from these panels that multiple velocity components are present,
especially in the molecular gas, and that the moment analysis may indeed
not be quite adequate in the CNR. The individual diagrams in Figs.~11
and 12 show velocity as a function of offset from the center in the
horizontal direction, either along the kinematic minor axis of the
galaxy, or along a line parallel to the minor axis but offset from it in
the vertical direction. The central panel in Figs.~11 and 12 is the
minor axis plot, the others are offset by \secd 1.5 (CO) and \secd 0.4
(\ha; not all panels are shown in Fig.~12), corresponding to roughly
half a resolution element in each case. The position angle of the galaxy
was taken as 153\deg\ (Knapen et al. 1993), thus that of the minor axis
is 63\deg\ (the latter is indicated in Fig.~4). As noted by Knapen et
al. (1993) in their \hi\ study, the bar of M100 is almost perfectly
placed for kinematic studies, its major axis being 110\deg, or
45\deg$\pm$2\deg\ from the kinematic major axis of the galaxy!

In the CO data (Fig.~11) four interesting components show up
symmetrically in the position-velocity diagrams, north and south of
the minor axis. Two of these (labeled AN and AS in Fig.~11) are
visible at RA offsets $\sim-8\sec$ (North) and $\sim+8\sec$ (S), with
excess velocities with respect to $v_{\rm sys}$ of $\sim+25\kms$ (N)
and $\sim-25\kms$ (S). We identify these components as the density
wave streaming motions near the spiral armlets, seen before in the
velocity field. Their offsets, both in RA and in velocity, strongly
support this interpretation, as well as their symmetric occurrence in
the series of position-velocity diagrams. The positions indicated by
letters in Fig.~11 correspond to positions in Fig.~4, the total
intensity CO map, of RA=\hmsd 12h22m54.24s, dec=\dms 15d49m22s (AN),
and \hmsd 12h22m55.53s, \dms 15d49m18s (AS). These positions must be
used as indications only, because the components can be seen to extent
to different offsets from the minor axis (Fig.~11) and because the
beam size in CO makes it more difficult to pinpoint positions in
Fig.~4.

The second set of components (labeled BN and BS in Fig.~12) has RA
offsets of $\sim+3\sec$ (N) and $\sim-3\sec$ (S), and excess
velocities of $\sim+70\kms$ (N) and $\sim-70\kms$ (S). Indicative
positions in Fig.~4 would be RA= \hmsd 12h22m54.90s, dec=\dms
15d49m26s (BN), and \hmsd 12h22m54.90s, \dms 15d49m13s (BS). These
components result from the gas streaming along the inner bar-like
feature, as again indicated by the symmetric offsets in both position
and velocity, and by qualitative and quantitative comparison with the
velocity field. This kinematically observed gas streaming confirms the
existence of the inner bar component, so prominently seen in our NIR
imaging and dynamical modeling (Knapen et al. 1995a,b).

The \ha\ position-velocity diagrams, shown in Fig.~12 in an analogous
manner to those for CO described above, albeit at higher spatial
resolution, are much harder to interpret.  The main reason for this is
the intrinsically very high velocity dispersion of powerful \hii\
regions (Rozas et al.  1998; Zurita, Rozas \& Beckman, in preparation),
which account for most of the \ha\ emission in the CNR.  This effect can
be easily seen in Fig.~12, and to a certain extent masks the effects of
gas streaming.  Even so, the main trends described before on the basis
of the CO position-velocity diagrams, and exemplified by the components
AN, AS, BN and BS, can be recognized in Fig.~12, and thus reinforce the
conclusions drawn above. 

It is interesting to compare the bar-induced deviations described here
for the small scales with those due to the large-scale bar, or Fig.~12
of the present paper with fig.~9 of Knapen et al. (1993). The latter
are similar position-velocity diagrams along and parallel to the minor
axis of M100, but show the \hi\ kinematics on the scale of the large
bar, which extends some 4~kpc in radius along a position angle of
$\sim110\deg$ (the position angle of the inner part of the bar is
identical, Knapen et al. 1995b). The similarity between the gas
components labeled BN and BS here, and the \hi\ components with
velocities symmetrically offset from $v_{\rm sys}$ is striking. The
\hi\ behavior was interpreted by Knapen et al. (1993) as streaming
along the large-scale bar of the galaxy.  The CO velocities similarly
represent the streaming along the inner part of the bar.  These
kinematic observations indicate the geometrical similarity of the gas
motions on the large and small scales.

\section{Numerical modeling}

In order to compare the results of our kinematic observations to our
modeling of the central region of M100 in a more quantitative way,
we have analyzed the results of the gas motion in our 3D numerical
model for stars and gas, in a similar fashion to our analysis of the 
observational data (using the Q1 model of Knapen et al. 1995b).  We
used the model parameters at a representative time $\tau=20$, when a
sufficient amount of gas has accumulated in the CNR due to the gas inflow
along the bar from larger radii, and the overall evolution has reached a
quasi steady state. The  morphology and gas distribution
in the model stay similar to that in the CNR of M100, for  a
few $\times 10^8$~yrs, and we use only the most robust morphological 
features for comparison with observations.
Non-linear orbit analysis (Knapen et al. 1995b) shows that two ILRs
exist at this stage, the outer ILR at about 20\sec\ and the inner ILR
at about $7\sec-8\sec$ (the {\it erroneous} linear model gives 27\sec
and 3\sec, respectively). This manifests itself by an appearance of the
$x_2$ orbits, oriented perpendicularly to the major bar axis. The gas
within the corotation radius is losing its angular support due to the
gravitational torques from the bar and moves inwards toward the CNR. In
this process the gas orbits gradually change their orientation from
being aligned with the bar ($x_1$ orbits) at larger radii to $x_2$
orbits between the ILRs. The two spiral arms coming into the CNR
represent the present positions of grand-design shocks in the gas. 
The incomplete nuclear ring forms where the gas settles down on more
circular $x_2$ orbits. Inside the inner ILR, the gas motions gradually
orient themselves again with the bar major axis. Knapen et al. (1995a,b)
find that the most intensive SF happens at the inner edge of the
nuclear ring, i.e., in the vicinity of the inner ILR.

We produced a velocity field from the model output, which is shown in
Fig.~13. The model was first rotated to the observed orientation based
on the position of the stellar bar.  The line-of-sight velocity at the
mid-plane was then determined for each pixel using the SPH kernel.
The smoothing length used was determined by the radius of a sphere
containing a minimum of 96 gas particles and restricted to be no
smaller than the adopted pixel size of $100\times100\,{\rm pc}^2$.  In
this way the velocity field evaluation is consistent with the
hydrodynamical method used in the model, is not sensitive to the pixel
size, and has  approximately the same accuracy everywhere.  The
resolution varies from 100\,pc at the center to 550\,pc at the edges
of the frame. The corresponding gray-scale map of shock dissipation 
in the CNR was published by Knapen et al. (1995b, their fig.~13, 
frame $\tau=20.0$).

In comparing the model velocity field with the observational data, a
few differences in characteristics must be kept in mind. The spatial
resolution in the SPH model map is not constant, as explained above.  We
smoothed our model results in such a way that the resolution in the map
is nowhere higher than that of the \ha\ data. For our  rotation curve
comparison (see below) we used the estimated  distance to M100
by Freedman et al. (1994).

A qualitative analysis of the model velocity field (Fig.~13) shows that
it agrees remarkably well with the observations. The general shape of
the velocity field is similar, with strongly rising circular velocities
in the central region, and two regions of maximum velocity indicated by
closed contours along the major axis, indicating a maximum in the
rotation curve. The two main types of non-circular motion identified in
the observations show up prominently and at the expected positions in
the model, namely those resulting from streaming of gas along the inner
part of the bar (S-shaped isovelocity contours) and from density wave
streaming motions near the spiral arms. This indicates that the model
reproduces the observed morphology of the central region of M100 and
its kinematic structure.

As with the CO observational data, we derived the synthetic gas (i.e.,
SPH)  rotation curve from our model. Fig.~14 shows the full resolution
curve, and an additional curve obtained from a velocity field  smoothed
in such a way that the highest model resolution, in the center,
corresponds to that of the CO. We also reproduce the CO rotation curve
made while excluding all points $>15\deg$ from the major axis, as shown
in Fig.~8  (abbreviated CO-75). Due to the variable resolution in the
model, the CO-75 curve should be compared to the smoothed model curve for $<
10\sec$, and to the  unsmoothed one for $>10\sec$.

Comparison between the synthetic and observed gas rotation curves 
shows that the
general shape of the curves is very similar: both curves rise steeply,
show a local maximum between $5\sec$ and $10\sec$, and fall off after
that. The sharp initial rise in the rotation is a clear signature of a
central mass concentration in the core of M100. Numerical simulations
reveal the expected trend, i.e., gas inflow towards the CNR and its
dominant role in re-shaping the gravitational potential there. Whether
this evolution is accompanied by a continuous buildup of stellar bulge
is an interesting question, but clearly beyond the scope of this
work. We only comment that SF processes can affect the evolution of
the CNR profoundly, and must be taken into account in any reasonable
model. In particular, as shown by Knapen et al. (1995b), this
evolution results in the gas being pushed inwards, across the inner
ILR. Both modeling and observations display a sharp decrease in the SF
within the inner $7\sec-8\sec$, i.e., close to the inner ILR. Gas
``filtering'' across the resonance region in the model occurs on a
characteristic timescale of about $10^9$ yrs.
  
\begin{figure*}
\epsfxsize=17cm \epsfbox{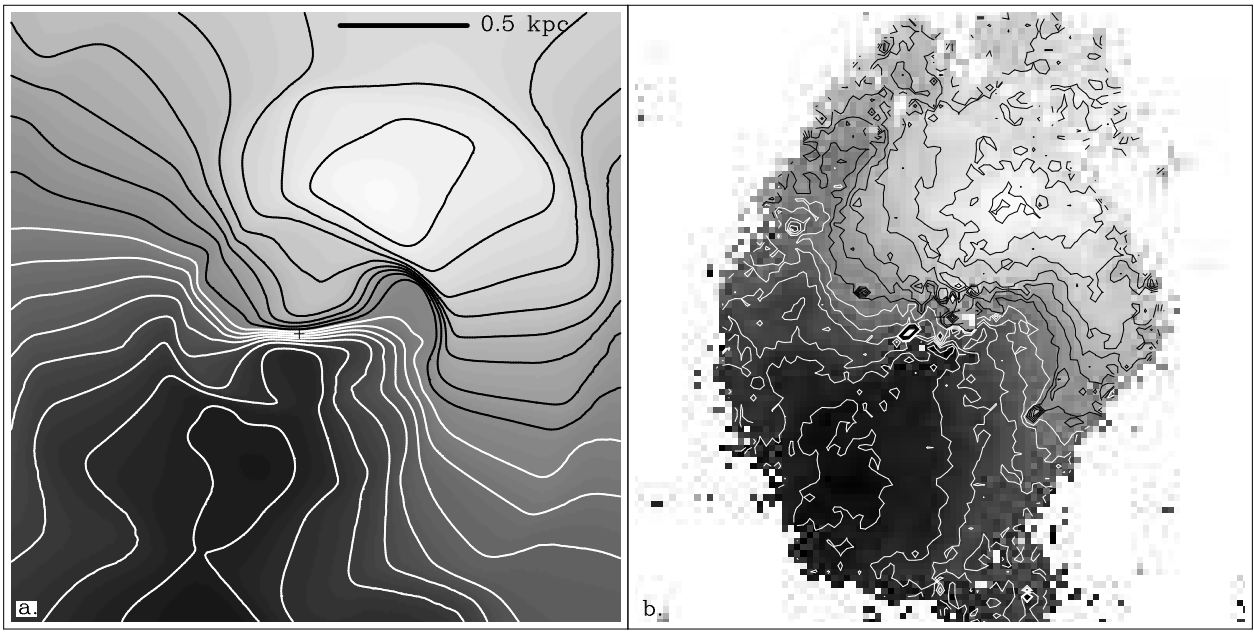}
\caption{Left panel ({\it a.}): Gas velocity field as derived from the
numerical model of Knapen et al. (1995b). Contour separation is 15\kms, and
the scale is indicated in the top right hand corner. At the distance of
M100, the total size of the region shown here would correspond to some
29\sec. N is up, E to the right. The position angle of the major axis is
as in M100. Resolution is highest in the center but lower than 100~pc
everywhere, making it comparable to our \ha\ data. Right panel ({\it
b.}): For comparison, the \ha\ velocity field of Fig.~2 is shown at
the same scale as the model velocity field. Contour separation and
orientation are also equal.}
\end{figure*}

\begin{figure}
\epsfxsize=8cm \epsfbox{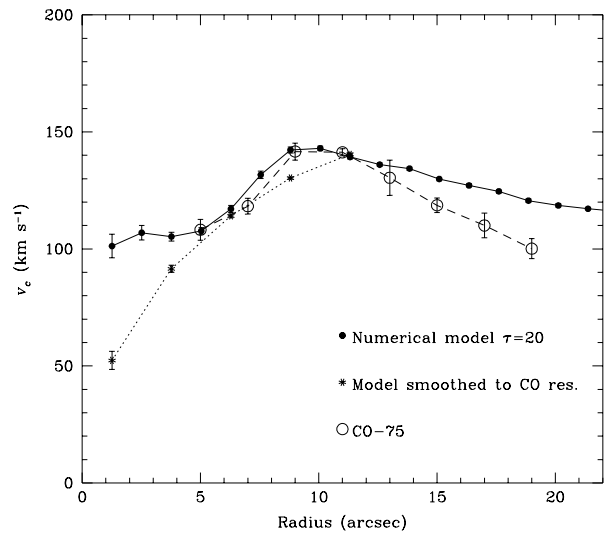}
\caption{Gas rotation curve as derived from the numerical velocity field of
Fig.~13 (filled dots, drawn line). Also shown is the SPH curve smoothed
to the CO resolution (stars, dotted line). 
CO-75 (see Fig.~8) is shown for comparison. Due to variable resolution,
in the model, the CO-75 curve should be compared to smoothed CO for 
$< 10\sec$ and to unsmoothed CO for $> 10\sec$.}
\end{figure}

\section{Discussion and interpretation}

\subsection{Double or single bar in M100?}

There has been some debate in the recent literature as to whether M100
has only one bar of which we observe the inner and outer parts, which
have the same pattern speeds (Knapen et al. 1995a; Wada et al. 1998), or
whether there are two separate bar systems, with different pattern
speeds, which are aligned by chance to within a few degrees in major
axis position angle (Garc\'\i a-Burillo et al. 1998). Double or nested
bars are quite common (e.g., Friedli et al. 1996; Jungwiert, Combes \&
Axon 1997), and many known cases host a Seyfert or otherwise active
nucleus, although precise statistics are not available. Systematic
studies of such systems at high resolution are long overdue because
nested or double bar systems are of interest to inflow studies,
especially with regard to the fueling of nuclear activity (Shlosman,
Begelman \& Frank 1990).

Although aligned nested bars are expected statistically, they would not
be obvious when observing barred galaxies. Nested bars are usually found
by studying (deprojected) isophotes. An inner bar with a different
pattern speed (co-rotating or counter-rotating) than the primary bar
will have different gas and stellar kinematics.  This includes
dissimilar rotation curves and dispersion velocities, gas/dust and star
forming region distributions, and other dynamical
characteristics. Evolutionary patterns of the CNRs should differ in all
these cases, which are possible to discriminate observationally.

In M100, the observed and modeled morphologies of the CNR, including the
distribution of star formation, allow limited freedom for
interpretation of gas and stellar dynamics within the central kpc.  The
remarkable agreement between both ellipticity and major axis position
angles of the inner and outer parts of the bar, the distribution of star
formation, the gas kinematics and especially the twisting of the NIR
isophotes can be convincingly interpreted as evidence for only one bar,
dissected by the resonance region hosting nuclear ring. 

The following three points support our interpretation of a double ILR in
the CNR of M100 (Knapen et al 1995b). First is the observed strong
offset of the leading dust lanes from the bar major axis. This offset is
directly related to the presence of the $x_2$ family of orbits
(Athanassoula 1992). In the absence of the ILR, these orbits disappear
and the dust lanes are centered along the bar axis. Nonlinear orbit
analysis displays part of the phase space occupied by these
orbits. Second is the twisting of $K$ isophotes (observed) and density
contours (modelled) backwards from the PA of the nuclear ring towards
the PA of the bar, both outside and inside the ring. Third, the overall
distribution of SF regions in the CNR, characterized by four main peaks
of SF, is explained by a global compression wave which is bar-driven and
which consists of two pair of arms, trailing and leading, interacting
off the bar major axis. The two main SF regions have been shown by the
model to lie at the caustic formed by the interaction of both pairs of
shocks, and two additional SF sites appear to be regions of gas
compression on the bar minor axis, so-called ``twin peaks''. The
simulations by Wada et al. (1998) are in broad agreement with these
findings.

\subsection{Star formation in the core region of M100}

It is obvious from \ha\ imaging (e.g. Arsenault et al. 1988; Pogge 1989;
Cepa \& Beckman 1990; Knapen et al. 1995a,b; Knapen 1998) that the CNR
of M100 is a site of strong massive star formation.  Knapen (1998) found
that luminosity function (LF) of the H~II regions in the CNR is very
flat. A flatter LF can in principle be due to an initial mass function
(IMF) skewed to enhanced SF at the high-mass end (Rozas 1996; Knapen
1998). However, a proposed change in the IMF slope must be regarded with
much suspicion even in the strongly star-forming CNR under
consideration, given the rather convincing evidence for a constant IMF
slope in an enormous range of environments (see Elmegreen 1999 for a
review).

A more realistic explanation of the flattened \hii\ region LF is an
observational one: large numbers of small \hii\ regions may be present,
thus assuring an ``underlying'' normally-shaped LF, but they cannot be
identified in an \ha\ image due to crowding effects. An alternative
effect, physically different, but leading to the same observational
result, is that many smaller \hii\ regions in a relatively small volume
coalesce to make up a much smaller number of much larger \hii\ regions.

Knapen et al. (1995a,b) discussed the $M/L$ ratio in the CNR,
especially as derived from a $K$-band image, and concluded on the
basis of population modeling and reasoning that there must be a
non-negligible contribution of dynamically young stars to the $K$-band
light. This is not surprising: population models
show that massive stars formed in a coeval or continuing
starburst evolve to a phase where their $K$ emission is strongly enhanced
after $0.5-1\times10^7$ years (e.g.  Leitherer \& Heckman
1995). Recently, several results on M100 have appeared in the
literature confirming this conclusion in a more quantitative way, and
due to the implications for several of the models for M100 that have
been published recently, we will briefly review those results here.

Wozniak et al. (1998) imaged the CNR of M100 with ISO at wavelengths of
6.75 and $15\mu$m. They found that the emission in the CNR is dominated
by three distinct regions, the nucleus, and the sites we identified as
K1 and K2 and confirmed to be active sites of star
formation by numerical modeling (Knapen et al. 1995a,b).  We postulated in
reference to optical and NIR morphology that K1 and K2, though different
in $K$ morphology and luminosity due to the presence of extincting dust,
are basically similar sites of strongly enhanced SF. This is well borne
out by the mid-IR imaging, much less sensitive to dust than imaging at
$2.2\mu$m. Ryder \& Knapen (1999) used a NIR color-color diagram and NIR
spectroscopy to study the stellar populations in K1 and K2. They
conclude from a comparison of the CO$_{\rm sp}$ versus Br$\gamma$ line
strengths with the models of Puxley, Doyon \& Ward (1997) that the stars
in K1 and K2 have ages between 20 and 25 Myrs. Finally, both Knapen et
al (1995b) and Wada et al (1998) found strong evidence for a
significantly changing $M/L_K$ across the CNR from a comparison of their
numerical modeling and NIR imaging, as discussed in more detail below.

\subsection{Comparison with other models}

The excellent agreement between the predictions of the dynamical model
of Knapen et al. (1995b) and the \ha\ and CO kinematics presented here
is strong evidence in favor of a single bar interpretation. This comes
in addition to the perfect alignment between the inner and outer parts
of the bar. The kinematic agreement is particularly remarkable since the
model was designed to elucidate the underlying physics of CNRs with
resonant structures, constrained only by the observed morphology of
M100, and constructed without any use of or reference to kinematic
observations of the region.

The model proposed by Garc\'\i a-Burillo et al. (1998), which claimed
two bars rotating at different angular speeds in M100, does not yield a
very detailed and close fit either to the morphology or the kinematics
of the CNR. In part, this is due to a number of simplifying assumptions
made. First, no gas self-gravity was included, whereas the gas mass
fraction in the CNR is at least 10\%. Sakamoto et al. (1995) measure the
overall gas fraction in the inner 18\sec\ of M100 to be in the range of
$\sim10\%-20\%$, and the spiral armlets can be expected to have even
more gas. For such a high gas fraction, evolution of models with and
without gas gravity is diverging (Shlosman 1999). Shaw et al. (1993)
presented an elegant proof that the twisting of the NIR isophotes cannot
be reproduced without gas self-gravity either. Second, the gas was
treated by means of a sticky particle code with an {\it ad hoc}
restitution coefficient. Shlosman \& Noguchi (1993) showed that gas
evolution, including radial inflow rates, depends strongly on this
parameter. Third, positions of CNR resonances were inferred from a
linear epicycle approximation. This is not suitable for M100-type bars
and gives erroneous positions for the ILRs, which require non-linear
orbit analysis (Athanassoula 1992; Knapen et al. 1995b). In addition,
the use of the 2D light distribution to infer the galactic potential so
close to the center, where the stellar disk thickness must be accounted
for, and where $M/L$ changes radially and azimuthally, further question
the approach taken by Garc\'\i a-Burillo et al.

Wada et al. (1998) compared interferometric CO observations by Sakamoto
et al. (1995) with a 2-D hydrodynamical model of the CNR in M100.  They
included self-gravity in the gas, and made a good attempt to follow the
variation in $M/L_K$ within the $K$-band image when deriving the
gravitational potential from their best fitting model, based on an
analytical description of the bar. The best fitting model invokes wide
variations in $M/L_K$ values, ranging from 1.2$M_\odot/L_\odot$ in the
older population zones, to 0.2 $M_\odot/L_\odot$ in what must be regions
of recent SF. This confirms a similar conclusion about the $M/L_K$
variation by Knapen et al. (1995b). The best model of Wada et al.  also
agrees well with the detailed morphology of the CNR containing a double
ILR, and with M100 having a single stellar bar with a pattern speed of
65\kms\,kpc$^{-1}$ (in excellent agreement with 68\kms\,kpc$^{-1}$ found
in Knapen et al.).

\section{Conclusions}

We compare high angular and velocity resolution two-dimensional
kinematic observations in the spectral lines of \ha\ and CO $J=1
\rightarrow 0$ of the circumnuclear starburst region in the barred
spiral galaxy M100 with kinematics derived from our numerical modeling
(Knapen et al. 1995b).  Our main results can be summarized as
follows: 

\begin{itemize}

\item We present Fabry-P\'erot kinematic mapping of the core region of
M100 at a fully sampled resolution of $\approx \secd 0.6$, and CO $J=1
\rightarrow 0$ interferometric observations. We present these data in
the form of channel maps, velocity fields, total intensity maps, and
selected position-velocity diagrams. From the \ha\ data, we derive a
rotation curve that rises rapidly in the central $\sim$140 pc, and is
roughly constant at the disk rotation velocity further out.

\item We compare CO and \ha\ rotation curves and argue that differences
between them are mainly due to the influence of streaming motions in the
gas, different for both tracers.  

\item We study in detail the characteristics of non-circular motions in
the core region by considering the CO and \ha\ velocity fields, residual
velocity fields after subtraction of the rotation curve, and sets of
position-velocity diagrams.  Non-circular motions are clearly identified
from both the \ha\ and CO data.By considering where the deviations from
circular motions occur in relation to morphological features and how
large they are, we can interpret them as the kinematic signatures of
spiral density wave streaming in the circumnuclear spiral arms, and as
gas streaming along the inner part of the bar.

\item We compare our observational data with a two-dimensional velocity
field and rotation curve derived from our 1995 dynamical model. The
observed and modeled kinematics show good qualitative and quantitative
agreement, for both the circular and non-circular kinematic
components. We interpret the results in terms of the prevailing orbits
in the combined gravitational potentials of stellar and gas
components. Our results are compatible with the presence of a global
density wave driven by a single, moderately strong stellar bar in
M100. Both the morphology and the kinematics of the CNR require the
presence of a double inner Lindblad Resonance in order to explain the
twisting of the near-infrared isophotes, distribution of star forming
region and the 2-D gas velocity field.

\item Finally, we review recent observational and modeling results on
the circumnuclear region in M100, and discuss the implications for bar
structure and gas dynamics in the core of M100 and other disk galaxies.

\end{itemize}

\acknowledgements 

We thank Fran\c{c}oise Combes for fruitful discussions and Seppo Laine
for comments on an earlier draft of this paper. Based on observations
obtained at the William Herschel Telescope, operated on the island of La
Palma by the Royal Greenwich Observatory in the Spanish Observatorio del
Roque de los Muchachos of the Instituto de Astrof\'\i sica de
Canarias. Financial support from the British Council and the Spanish
Acciones Integradas Programme, and from the Spanish DGES, Grant Nos.
PB94-1107 and PB97-0219, is acknowledged. I.S. is grateful for support
under NASA grants NAG5-3841, WKU-522762-98-06 and HST AR-07982.01-96A.

\end{document}